\documentclass[a4paper]{scrartcl}
\usepackage[T1]{fontenc}
\usepackage[latin1]{inputenc}
\usepackage{graphics}
\usepackage{graphicx}
\usepackage{verbatim}
\makeatletter

\providecommand{\LyX}{L\kern-.1667em\lower.25em\hbox{Y}\kern-.125emX\@}
\newcommand{\noun}[1]{\textsc{#1}}

\makeatother

\begin{document}

\titlehead{\texttt{}}

\vspace{2cm}
\title{Vector Meson Photoproduction at High-\protect\( t\protect \) and Comparison
to HERA Data}

\date{J.R. Forshaw and G. Poludniowski}

\maketitle
\vspace*{-0.9cm}

{\par\centering Department of Physics \& Astronomy\\
University of Manchester\\
Manchester M13 9PL.\\
U.K.\par}

\begin{abstract}
\bigskip{}
\noindent We explore QCD calculations for the process \( \gamma p\rightarrow VX \)
where \( V \) is a vector meson, in the region \( s\gg -t \) and \( -t\gg \Lambda _{QCD}^{2} \).
We compare our calculations for the \( J/\psi  \), \( \phi  \) and \( \rho  \)
mesons with data from the ZEUS Collaboration at HERA and demonstrate
that the BFKL approach is consistent with the data even for light mesons, whereas
the two-gluon exchange approach is inadequate. We also predict the differential
cross-sections for the \( \Upsilon  \) and \( \omega  \) for which no data
are currently available. 
\end{abstract}

\section{Introduction}

We study the process \( \gamma p\rightarrow VX \) (where \( V \) is a vector
meson: \( V= \) \( \rho  \), \( \omega  \), \( \phi  \), \( J/\psi  \)
or \( \Upsilon  \)) in the perturbative Regge limit, \( s\gg -t\gg \Lambda _{QCD}^{2} \).
The largeness of \( -t \) allows the application of the solutions of the non-forward
BFKL equation. We apply the analytic solutions to the cross-section arrived
at in \cite{A, B} in the case of a delta-function distribution for the meson
wavefunction. Specifically we apply this solution to the case of a real photon
which has been measured at HERA \cite{C,H1}. 

The cross-section for the helicity-flip process \( \gamma p\rightarrow V_{L}X \)
vanishes for a delta-function wavefunction and throughout this paper the cross-sections
pertain to the process \( \gamma p\rightarrow V_{T}X \) (where \( L \) and
\( T \) correspond to longitudinal and transverse polarization respectively).
This approximation is justifiable since the measured rate for longitudinal mesons
is indeed small \cite{C}. We make predictions and compare to experiment for
the mesons for which data are available, the heavy \( J/\psi  \) and the light
\( \rho  \) and \( \phi  \) (for which we might expect the delta-function
to be a poor approximation). We show that the data for all the mesons can be
well described by the BFKL approach. We then demonstrate that the two-gluon
exchange model is incapable of adequately describing the data. Finally we make
predictions for the \( \Upsilon  \) and \( \omega  \).

The high momentum exchange allows us to factorize the cross-section into the
usual product of the parton distribution functions and the parton level cross-section:

\begin{equation}
\frac{d\sigma (\gamma p\rightarrow VX)}{dtdx}=\biggl (\sum _{f}[q_{f}(x,t)+\bar{q}_{f}(x,t)]\biggr )\frac{d\sigma (\gamma q\rightarrow Vq)}{dt}+G(x,t)\frac{d\sigma (\gamma g\rightarrow Vg)}{dt},
\end{equation}
 where \( G(x,t) \) and \( q_{f}(x,t) \) are the gluon and quark parton distribution
functions respectively and we sum over flavour, \( f \). For a large separation
in rapidity between the parton and the vector meson we may write,

\begin{equation}
\frac{d\sigma (\gamma p\rightarrow VX)}{dtdx}=\biggl (\frac{4N_{c}^{4}}{(N_{c}^{2}-1)^{2}}G(x,t)+\sum _{f}[q_{f}(x,t)+\bar{q}_{f}(x,t)]\biggr )\frac{d\sigma (\gamma q\rightarrow Vq)}{dt},
\end{equation}
where we define the parton level amplitude by

\begin{equation}
\frac{d\sigma (\gamma q\rightarrow Vq)}{dt}\equiv \frac{\pi }{4t^{4}}|\mathcal{F}(s^{\prime},t)|^{2},
\end{equation}

where $s$ is the centre-of-mass energy squared of the photon-proton system and $s^{\prime}=xs$ is that of the photon-parton system.

\section{The Two Gluon and BFKL Amplitudes}

We consider two types of colour singlet exchange: two-gluon and BFKL. The two-gluon
amplitude can be expressed to leading order in \( s \) in the impact factor
representation:

\begin{equation}
\mathcal{F}_{Born}(s^{\prime},t)=2\pi t^{2}\int \frac{d^{2}k_{\perp }}{(2\pi )^{2}}\frac{\mathcal{I}_{\gamma V}\, \mathcal{I}_{qq}}{k_{\perp }^{2}(k_{\perp }-Q_{\perp })^{2}},
\end{equation}
where \( \mathcal{I}_{\gamma V} \) and \( \mathcal{I}_{qq} \) are the impact
factors associated with the couplings of the two gluons to the external particles.
Vectors with the subscript \( \perp  \) are two dimensional transverse momenta
and \( Q_{\perp }^{2}=-t \). With the above definitions the impact factor for
\( q\rightarrow q \) is given by

\begin{equation}
\mathcal{I}_{qq}=\frac{\delta _{ab}}{N_{c}}\alpha _{s}
\end{equation}

The \( \gamma \rightarrow V_{T} \) impact factor, assuming a delta-function
form for the meson wavefunction, is

\begin{eqnarray}
\mathcal{I}_{\gamma V}=\mathcal{C}\, \alpha _{s}\frac{\delta _{ab}}{2N_{c}}\Biggl (\frac{4}{Q_{\perp }^{2}+M_{V}^{2}}-\frac{1}{(k_{\perp }-Q_{\perp }/2)^{2}+M_{V}^{2}/4}\Biggr ), & \label{con} 
\end{eqnarray}
 where,

\begin{eqnarray}
\mathcal{C}^{2}=3\Gamma ^{V}_{e^{+}e^{-}}M_{V}^{3}/\alpha _{em}.
\end{eqnarray}
 \( M_{V} \) is the mass of the vector meson and \( \Gamma ^{V}_{ee} \) is
the electronic decay width of the meson. Using these impact factors we obtain
the two-gluon amplitude as a function of one dimensionless parameter, \( \tau =-t/M_{V}^{2} \):

\begin{equation}
\mathcal{F}_{Born}=\mathcal{C}\, \mathcal{I}_{qq}^{2}\biggl (\frac{4\tau ^{2}}{1-\tau ^{2}}\biggr )\ln \biggl (\frac{(1+\tau )^{2}}{4\tau }\biggr ),
\end{equation}

which has the mathematical property $\mathcal{F}_{Born}\rightarrow 0$ as $\tau\rightarrow 1$.

The BFKL amplitude, in the leading logarithm approximation (LLA), is given by
\cite{E},

\begin{equation}
\label{BFKLa}
\mathcal{F}_{BFKL}(s^{\prime},t)=\frac{t^{2}}{(2\pi )^{3}}\int d\nu \frac{\nu ^{2}}{(\nu ^{2}+1/4)^{2}}e^{\chi (\nu )z}I_{\nu }^{q^{\ast }}(Q_{\perp })I^{\gamma V}_{\nu }(Q_{\perp }),
\end{equation}
 where

\begin{equation}
\chi (\nu )=4\mathcal{R}e\biggl (\psi (1)-\psi \bigg (\frac{1}{2}+i\nu \bigg )\biggr )
\end{equation}
 and 
\begin{equation}
\label{zdef}
z=\frac{3\alpha _{s}}{2\pi }\ln \biggl (\frac{s^{\prime}}{\Lambda ^{2}}\biggr ).
\end{equation}
 In LLA, \( \Lambda  \) is arbitrary (it need only be small compared to \( \surd s^{\prime} \))
and \( \alpha _{s} \) is a constant. The impact factors are used, in conjunction
with the prescription of \cite{F}, to give

\begin{eqnarray}
I_{\nu }^{A}(Q_{\perp }) & = & \int \frac{d^{2}k_{\perp }}{(2\pi )^{2}}\, \mathcal{I}_{A}(k_{\perp },Q_{\perp })\int d^{2}\rho _{1}d^{2}\rho _{2}\\
 & \times  & \biggl [\left( \frac{(\rho _{1}-\rho _{2})^{2}}{\rho _{1}^{2}\rho _{2}^{2}}\right) ^{1/2+i\nu }-\left( \frac{1}{\rho _{1}^{2}}\right) ^{1/2+i\nu }-\left( \frac{1}{\rho _{2}^{2}}\right) ^{1/2+i\nu }\biggr ]e^{ik_{\perp }\cdot \rho _{1}+i(Q_{\perp }-k_{\perp })\cdot \rho _{2}},\nonumber 
\end{eqnarray}
In the case of coupling to a colourless state only the first term in the square
bracket survives since \( \mathcal{I}_{A}(k_{\perp },Q_{\perp }=k_{\perp })=\mathcal{I}_{A}(k_{\perp }=0,Q_{\perp })=0 \)
in this case. After some work, one obtains \cite{B}:

\texttt{
\begin{eqnarray}
I^{q}_{\nu }(Q_{\perp }) & = & -4\pi \mathcal{I}_{qq\, }2^{-2i\nu }|Q_{\perp }|^{-1+2i\nu }\frac{\Gamma (\frac{1}{2}-i\nu )}{\Gamma (\frac{1}{2}+i\nu )}\label{Iq} \\
I_{q}^{V}(Q_{\perp }) & = & -\mathcal{C}\, \mathcal{I}_{qq}\frac{\pi ^{2}}{Q_{\perp }^{3}}\frac{\Gamma (1/2-i\nu )}{\Gamma (1/2+i\nu )}\biggl (\frac{Q_{\perp }^{2}}{4}\biggr )^{i\nu }\int _{1/2-i\infty }^{1/2+i\infty }\frac{du}{2\pi i}\biggl (\frac{Q_{\perp }}{M_{V}/2}\biggr )^{1+2u}2^{2(1-u)}\\
 &  & \frac{\Gamma (1-u-i\nu )\Gamma (1-u+i\nu )\Gamma ^{2}(1/2+u)}{\Gamma (1/2+u/2-i\nu /2)\Gamma (1-u/2-i\nu /2)\Gamma (1-u/2+i\nu /2)\Gamma (1/2+u/2+i\nu /2)}.\nonumber \label{IV} 
\end{eqnarray}
}Putting these into (\ref{BFKLa}) we obtain,

\begin{eqnarray}
\mathcal{F}_{BFKL}(s^{\prime},t) & = & 4\, \mathcal{C}\, \mathcal{I}^{2}_{qq}\int d\nu \frac{\nu ^{2}}{(\nu ^{2}+1/4)^{2}}e^{\chi (\nu )z}\int _{1/2-i\infty }^{1/2+i\infty }\frac{du}{2\pi i}\tau ^{1/2+u}\label{result} \\
 &  & \hspace *{-2cm}\frac{\Gamma ^{2}(1/2+u)\Gamma (1-u-i\nu )\Gamma (1-u+i\nu )}{\Gamma (1/2+u/2-i\nu /2)\Gamma (1-u/2-i\nu /2)\Gamma (1-u/2+i\nu /2)\Gamma (1/2+u/2+i\nu /2)}.\nonumber 
\end{eqnarray}

\section{Results and Discussion}

Using the full numerical calculation for the amplitude we convoluted the partonic
cross-section for the mesons with the parton density functions of the proton\footnote{%
GRV-98 LO \cite{pdf}.
}, integrating over \( x \) in the region \( 0.01<x<1 \) (for which the ZEUS
data are quoted). We obtained fits to the ZEUS data in terms of three free parameters \cite{C}\footnote{Note that the differential cross-sections for the three mesons have an overall normalization uncertainty of approximately 10\% (which cancels in the ratios of cross-sections). This normalization uncertainty does not significantly affect our results.}. One parameter
is \( \alpha _{s} \) and the others appear in the denominator of the logarithm
defining the energy variable \( z \), i.e. we take \( \Lambda ^{2} = \beta M_{V}^{2}+\gamma |t| \).

Initially we tried the simple prescription of \cite{A, B} where \( \beta = \gamma =1 \).
The \( J/\psi  \) has been fitted with \( \alpha _{s}=0.20 \) for this parameterization
previously \cite{zp}\footnote{%
H1 has recently performed a similar fit where they found \( \alpha _{s} = 0.22 \)
\cite{H1}.
} and as expected this gave a \( \chi ^{2}/\mathrm{dof}<1 \), however these
parameters gave very poor fits to the \( \rho  \) and \( \phi  \). When we
varied all three parameters independently we found that optimum fits took small
values of \( \gamma  \) and in fact we were able to put this parameter to zero.

\begin{center} 
\begin{figure}
\vspace{0.5cm}
{\par\centering \resizebox*{9cm}{8cm}{\includegraphics{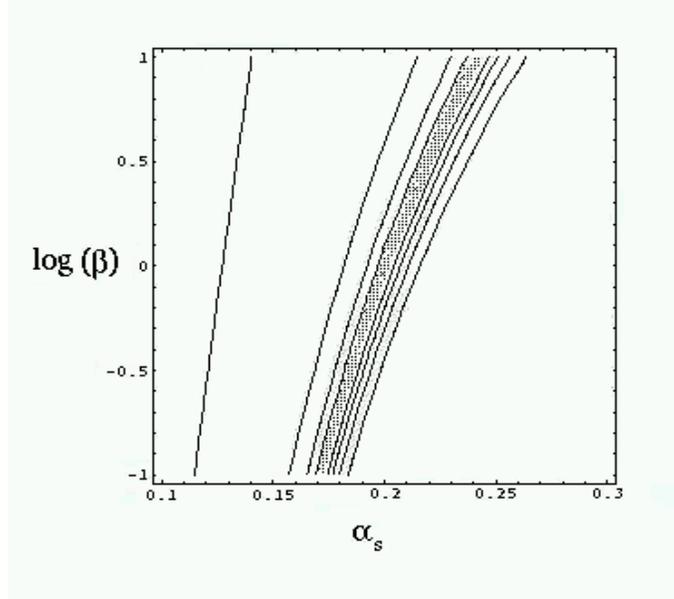}} \par}

\caption{Contour plot of $\chi^{2}/$dof where $\chi^{2}$ is that of the $J/\psi, \,\phi,\,\rho$ mesons combined. From the inside heading outwards the contours are those of $\chi^{2}/dof=1,\,3,\,9,\,27$, with the shaded region corresponding to $\chi^{2}/$dof$\le 1$.}
\label{contours}
\end{figure}

\end{center}

Fig. \ref{contours} shows contours of constant $\chi^{2}$. The contours continue beyond the region $0.1<\beta<10$ without closing. Further imposing the requirement that $s^{\prime}\ge 10\,\beta M_{V}^{2}$ as a sensible criterion for the dominance of leading logs, a simultaneous fit for the three mesons rules out the region of $\alpha_{s}>0.2$ in Fig. \ref{contours}. In addition, sub-asymptotic leading log contributions become increasingly important as we increase the strong coupling to $\alpha_{s}>0.20$ \cite{EMP}. Fig. \ref{contours} suggests that we can get acceptable fits for $\alpha<0.17$; while this is true we are then forced to accept unnaturally small values of $\beta$. We note that this range of $0.17 \le\alpha_{s}\le 0.20$ is consistent with the value extracted from the Tevatron ``gaps between jets'' data \cite{CFL}. 

It should be observed that the acceptable fits lie in a narrow valley in parameter space. It turns out that the relation,  

\begin{equation}
\alpha_{s}(\beta)=\frac{\alpha_{c}}{1-\alpha_{c}\,d\, \ln \beta}
\end{equation}

with $d=0.358$ gives a good approximation of the flow of $\alpha_{s}$ with $\beta$ for a constant $\chi^{2}$. The width of the $\chi^{2}/$dof$\le 1$ valley corresonds to a band $0.197\le \alpha_{c} \le 0.204$, with the minimum at approximately $\alpha_{c}=0.200$. 

\begin{figure}
\pagebreak
{\par\centering {\includegraphics[height=10cm,width=6cm,angle=-90]{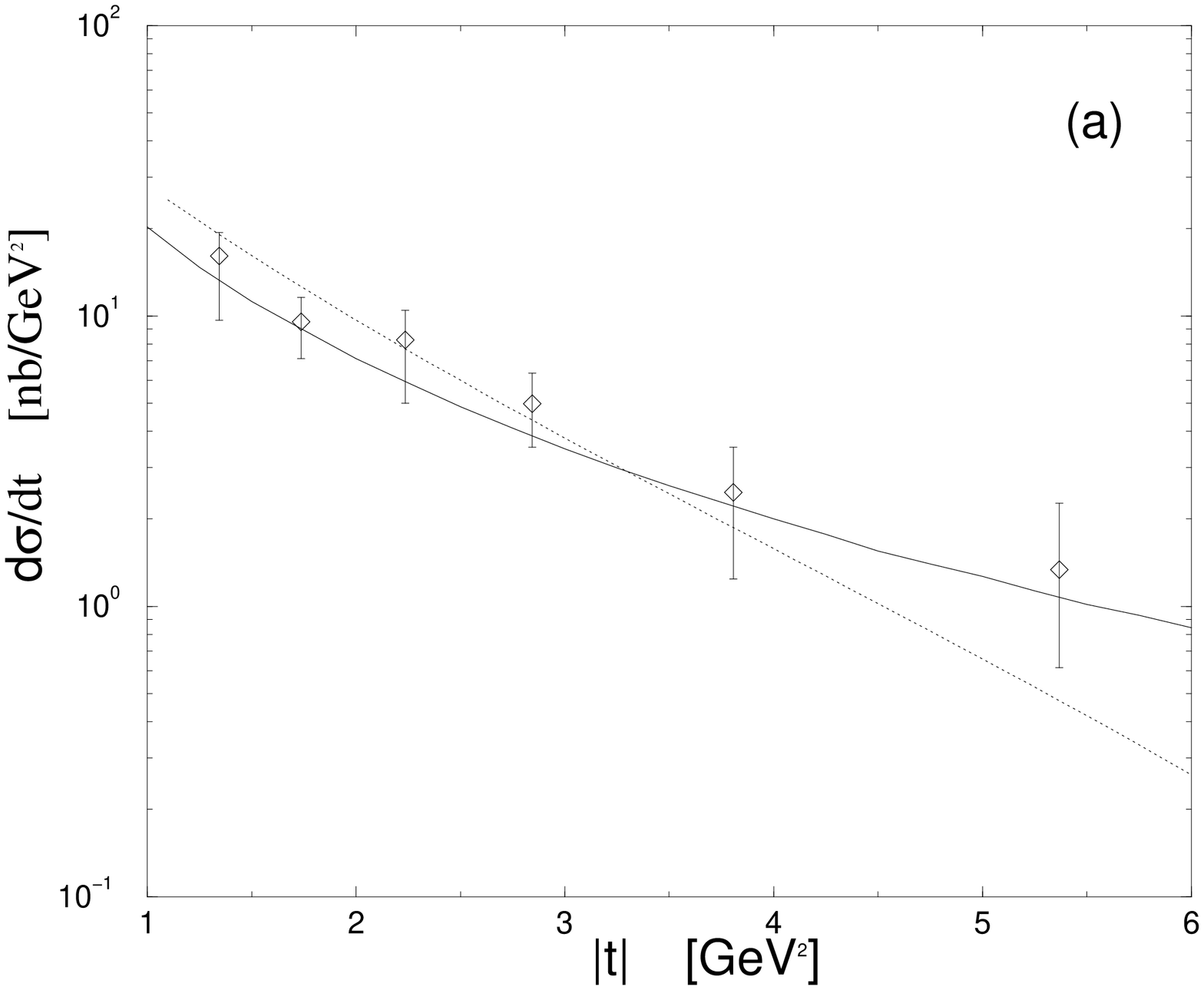}} \par}
{\par\centering {\includegraphics[height=10cm,width=6cm,angle=-90]{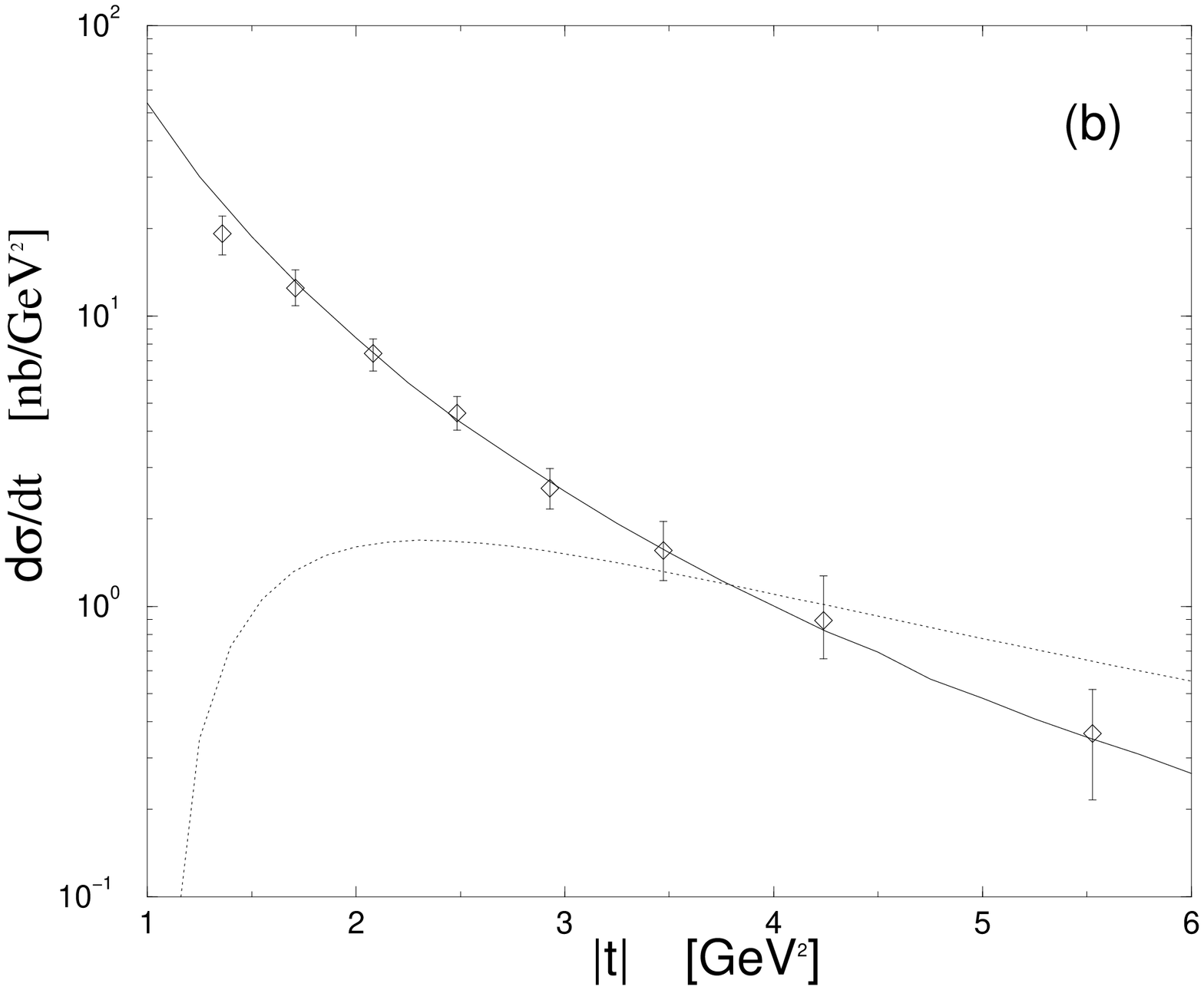}} \par}
{\par\centering {\includegraphics[height=10cm,width=6cm,angle=-90]{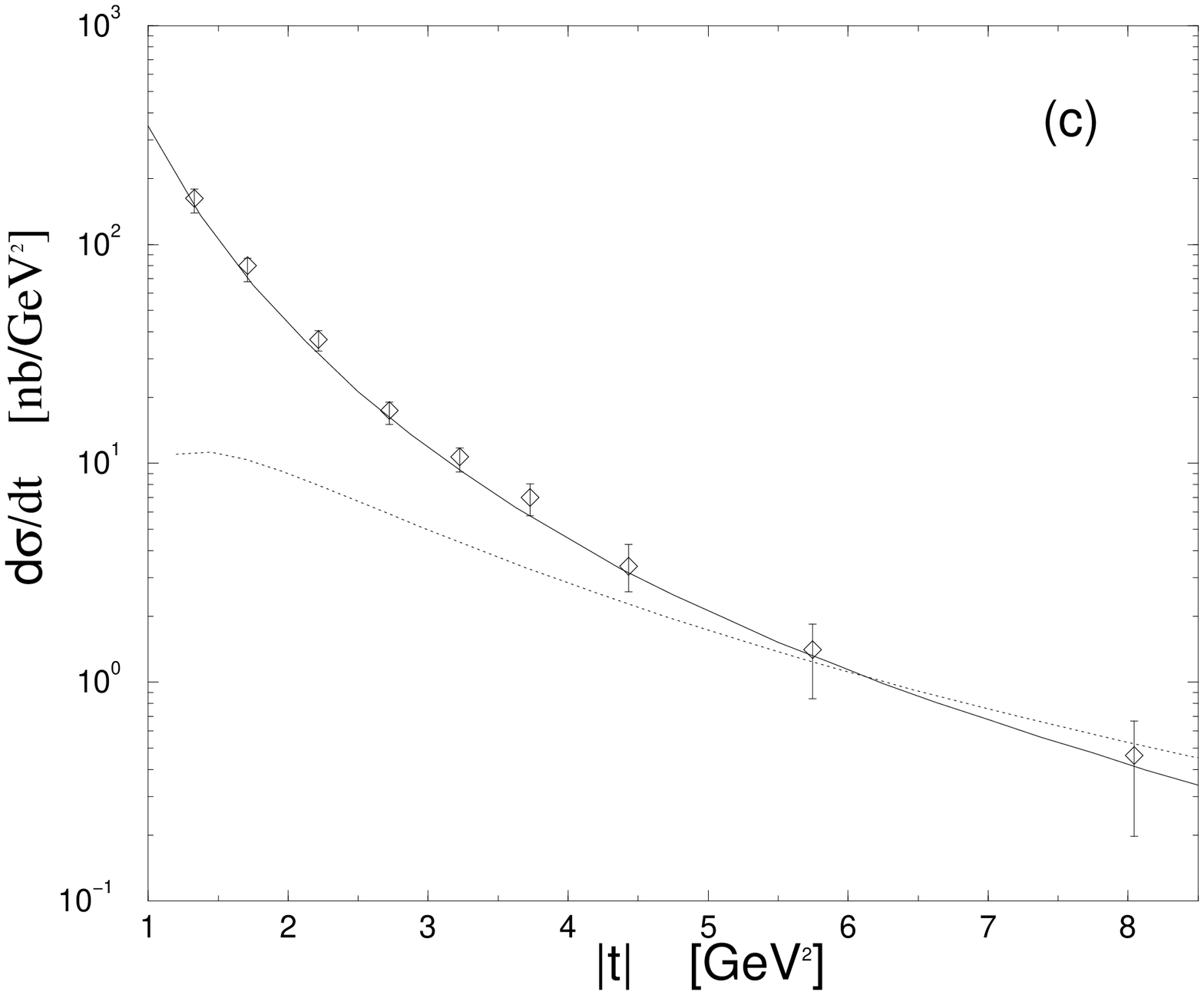}} \par}

\caption{Comparison of BFKL (solid line) and two-gluon (dotted line) calculations compared
to ZEUS data \cite{C} for (a) \protect\( J/\Psi \protect \), (b) \protect\( \phi \protect \)
and (c) \protect\( \rho \protect \) meson production \cite{C}. The BFKL curves
correspond to the fit ($\alpha_{s}$, $\beta$, $\gamma$) = (0.20, 1.0, 0.0), described in the text. The two-gluon curves are obtained
by optimising \protect\( \alpha _{s}\protect \) in each case, i.e. (a) \protect\( \alpha _{s}=0.40\protect \),
(b) \protect\( \alpha _{s}=0.35\protect \) , (c) \protect\( \alpha _{s}=0.27\protect \).}
\label{diffcross}
\end{figure}

Perhaps the most natural fit is ($\alpha_{s}$, $\beta$, $\gamma$) = (0.2, 1.0, 0.0). The BFKL differential cross-section predictions for the three mesons are shown in Fig. \ref{diffcross}, along with the ZEUS data \cite{C}.

Our results suggest that the largeness of \( -t \) does allow a perturbative
calculation to be performed however it is reasonable to ask whether it is necessary
to invoke the full machinery of BFKL. Hence we also show the two-gluon exchange
fits in Fig. \ref{diffcross}. The two-gluon calculation only has one free
parameter, \( \alpha _{s} \), and the differential cross-section has a fourth
order power dependence on it. It is apparent that this approximation provides a very
poor description of the light mesons, for which \( \tau \sim 1 \) over much
of the range of data. Note that the model starts to give the correct qualitative
behaviour away from \( \tau =1 \). Note in particular that \( \alpha _{s} \) has
been fitted for each meson separately and also the unnatural trend that \( \alpha _{s} \)
falls as \( M_{V} \) falls, which means that running the coupling only makes
the situation worse. The $J/\psi$ differential cross-section, to which we would expect our calculation to be most applicable can be fitted by the two-gluon exchange amplitude in the ZEUS $t$-range. For a plausible strong coupling $\alpha_{s}=0.4$ we obtain $\chi^{2}_{J/\psi}/$dof$=0.8$. It should be noted however that the two-gluon curve is already starting its dip towards zero at $|t|=M_{J/\psi}^{2}$, by the end of the ZEUS $t$ range. H1 data for this process reaching out to much higher $|t|$ should make it clear whether or not two-gluon exchange is sufficient to describe the $J/\psi$. 

\begin{figure}
{\par\centering {\includegraphics[height=10cm,width=6cm,angle=-90]{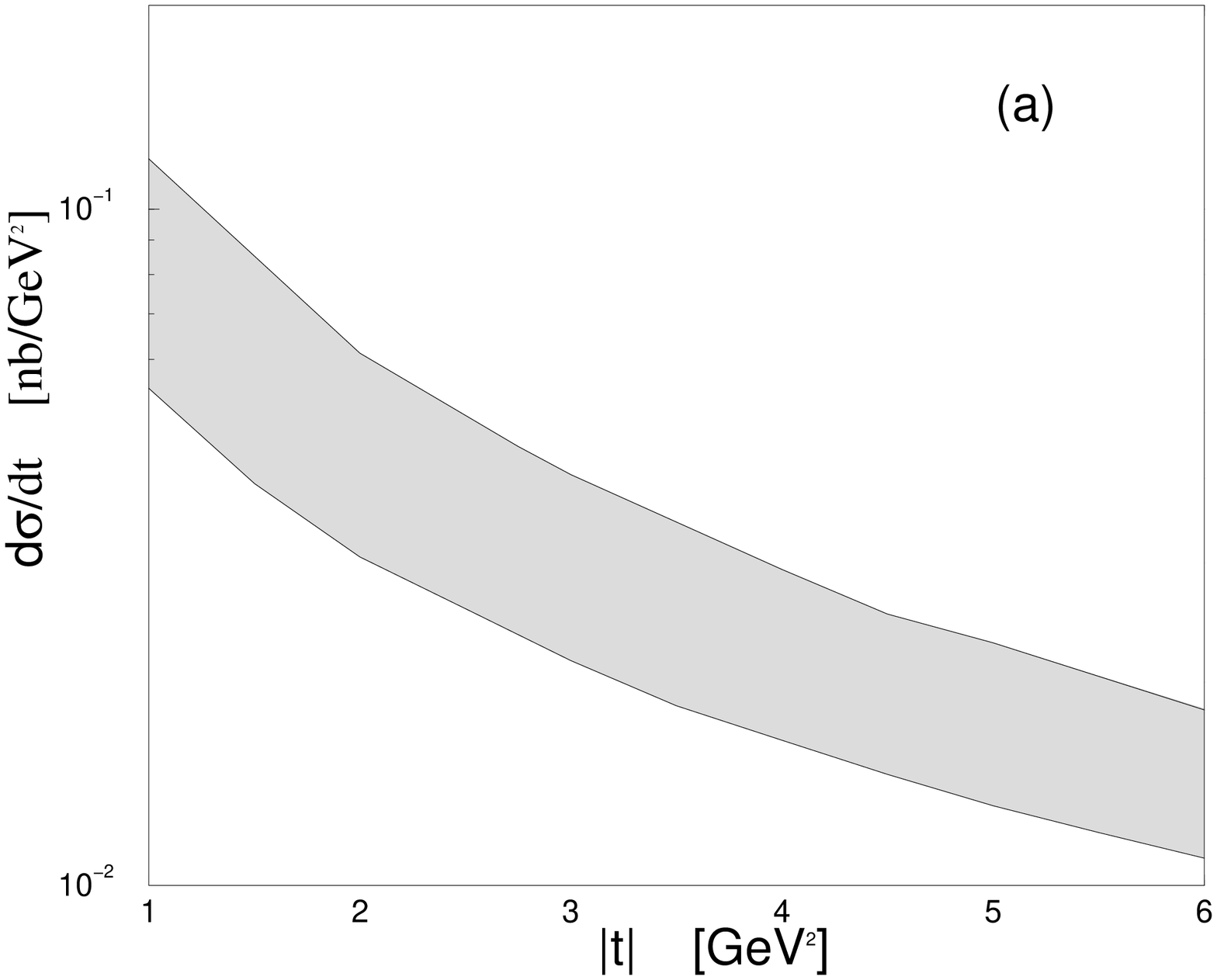}} \par}
{\par\centering {\includegraphics[height=10cm,width=6cm,angle=-90]{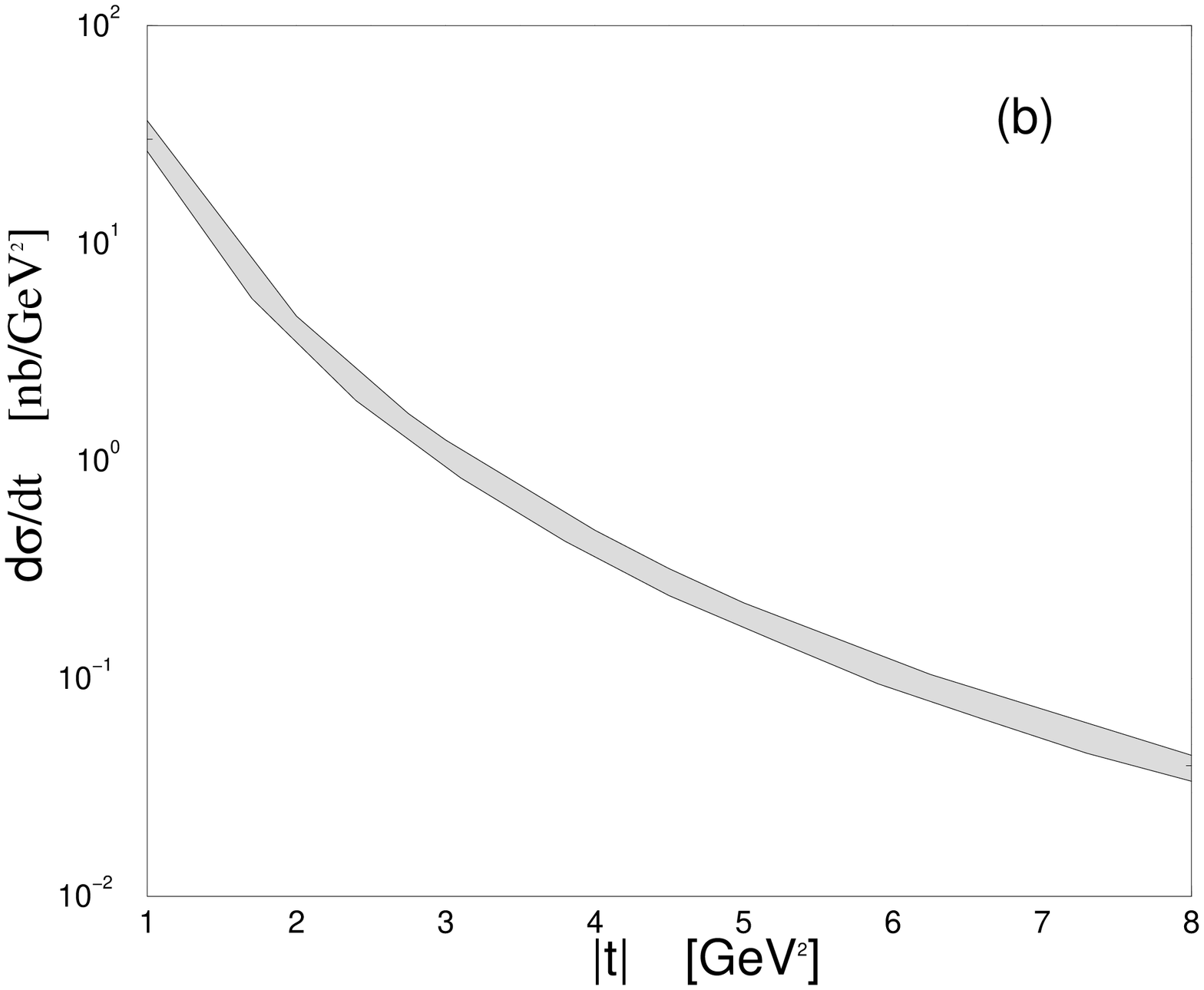}} \par}

\caption{Predictions for (a) \protect\( \Upsilon \protect \) and (b) \protect\( \omega \protect \)
meson production. The bounds come from varying $\alpha_{s}$ with $0.1\le \beta \le 1.0$ and keeping $\chi^{2}/$dof$\le 1$. The upper and lower bounds correspond to ($\alpha_{s}$, $\beta$) = (0.175, 0.1) and ($\alpha_{s}$, $\beta$) = (0.197, 1.0) respectively.}
\label{prediction}
\end{figure}

Fig. \ref{prediction} shows our predictions for the $\Upsilon$ and $\omega$ mesons. The bounds come from the uncertainty in the normalization of the LO BFKL solution, specifically from varying $\alpha_{s}$ while keeping $0.1 \le\beta\le 1.0$ and $\chi^{2}/$dof$\le 1.0$. The major part of the uncertainty is in normalisation; the shape being rather well predicted. The bounds for the \( \omega  \) are clearly much narrower than
for the \( \Upsilon  \). This is due to the fact that \( M_{\rho }\sim M_{\omega } \)
which means that \( d\sigma _{\omega }/dt\sim (\Gamma ^{\omega }_{ee}/\Gamma ^{\rho }_{ee})\cdot d\sigma _{\rho }/dt \).
This scaling relationship between the \( \rho  \) and \( \omega  \) is a model
independent feature and also true of the two-gluon model. It should be noted that as we approach the lower bound for the $\Upsilon$ in Fig. \ref{prediction} (a), we are extending the model beyond the leading log region (due to largeness of $\beta M_{\Upsilon}^{2}$), so this bound must be viewed with some caution. 

If subsequent measurements yield data lying outside these bounds it would indicate either a failure of the LLA BFKL formalism, or a need to refine our treatment of the meson wavefunction. If other light vector mesons,
such as the \( \omega  \), subsequently confirm our predictions it would suggest
that in some instances even mesons made of light quarks can act as if they consist
of two constituent quarks which share the meson's energy and momentum.

For all of the measured mesons, the experimental errors are about 10-20\% of the absolute
value of the differential cross-section. That we can fit the data suggests that
genuine higher order corrections and the corrections due to the meson wavefunction
may contribute no more than 10-20\%.

\section*{Final Remarks}

In \cite{I} a LLA BFKL calculation incorporating a more sophisticated meson
wavefunction than that used here is performed. It was concluded in \cite{C}
that this light meson calculation is incompatible with the available data. However,
the author of \cite{I} interprets the masses appearing in (\ref{con}) as current
masses (as ought to be appropriate for a truly perturbative calculation) which
justifies the neglect of the quark mass for light mesons, rendering \( |t| \)
the only relevant scale. By interpreting the quark mass as a constituent mass,
and assuming that the quark and anti-quark share the meson momentum, we have
demonstrated that the data can be understood provided the constituent mass is
taken to be the scale in \( \Lambda  \). Another point of deviation is in the
treatment of the strong coupling \( \alpha _{s} \). In \cite{I} the coupling
is a running coupling whilst we have considered a fixed coupling. The problem
with running the coupling it is that we do not really know how it runs. Of course
a complete description should incorporate a running coupling, but the work of
\cite{Brodsky} might be a hint that a fixed coupling may be appropriate for
the BFKL exchange. The work of \cite{I} has been developed in \cite{IKSS}
where it is proposed that a large contribution arising from \( q\bar{q} \)
fluctuations in a chiral-odd spin configuration should play an important role
in the region of the data. It remains to be seen if the simple model presented
here can be justified within this approach. 

Our results suggest that the two-gluon model is inadequate  at least for the light mesons.
The model predicts a dip at \( \tau =1 \) which is not present in the data
and this is the biggest obstacle to getting a decent fit. Corrections to the
two-gluon approximation do fill in the dip (see \cite{A} for more details on
how this occurs) and further study is called for to establish whether a more
sophisticated two-gluon or finite order (in $\alpha_{s}$) calculation might work. 

The model explored here has experimental tests to face in the future. If data
confirm our predictions for the \( \Upsilon  \) and \( \omega  \) it will
be impressive. Data on the process \( \gamma p\rightarrow \gamma X \),
for which theoretical calculations have already been performed \cite{X}, should
be obtained in the forseeable future. This process along with the ones considered
here will continue to provide an important test of the validity of BFKL dynamics.

\section*{Acknowledgement}

We thank the ZEUS collaboration for providing the data used in this paper and in particular we also thank James Crittenden, Brian Foster, Katarzyna Klimek, Yuji Yamazaki and Giuseppe Iacobucci for their help and advice.

\pagebreak
\pagebreak

\pagebreak 

\pagebreak

\end{document}